# Layer-by-layer growth of bilayer graphene single-crystals enabled by self-transmitting catalytic activity

Zhihong Zhang[1,2,#], Linwei Zhou[1,#], Zhaoxi Chen[3,4,#], Antonín Jaroš[5,6], Miroslav Kolíbal[5,6], Petr Bábor[5,6], Quanzhen Zhang[7], Changlin Yan[1], Ruixi Qiao[2], Qing Zhang[3], Teng Zhang[7], Wei Wei[8], Yi Cui[8], Jingsi Qiao[7], Liwei Liu[7], Lihong Bao[9,11], Haitao Yang[9,11], Zhihai Cheng[1], Yeliang Wang[7], Enge Wang[2], Zhi Liu[3,4], Marc Willinger[10], Hong-Jun Gao[9,11], Kaihui Liu[2,*], Zhu-Jun Wang[4,3,10,*], and Wei Ji[1,*]

[1] Beijing key laboratory of optoelectronic functional materials and micro-nano devices, Department of Physics, Renmin University of China, 100872, Beijing, China

[2] State Key Laboratory for Mesoscopic Physics and Frontiers Science Center for Nano-optoelectronics, School of Physics, Peking University, Beijing 100871, China.

[3] Center for Transformative Science, ShanghaiTech University, Shanghai 201210, China

[4] School of Physical Science and Technology, ShanghaiTech University, Shanghai 201210, China

[5] CEITEC BUT, Brno University of Technology, Purkyňova 123, 612 00 Brno, Czech Republic

[6] Institute of Physical Engineering, Brno University of Technology, Technická 2, 616 69 Brno, Czech Republic

[7] School of Information and Electronics, MIIT Key Laboratory for Low-dimensional Quantum Structure and Devices, Beijing Institute of Technology, Beijing 100081, China

[8] Vacuum Interconnected Nanotech Workstation, Suzhou Institute of Nano-Tech and Nano-Bionics, Chinese Academy of Sciences, Suzhou 215123, China

[9] Beijing National Laboratory for Condensed Matter Physics and Institute of Physics, Chinese Academy of Sciences, 100190, Beijing, China

[10] Scientific Center for Optical and Electron Microscopy, ETH Zurich, Otto-Stern-Weg 3, 8093 Zurich, Switzerland

[11] School of Physical Sciences, University of Chinese Academy of Sciences, Beijing 100039, China

**Direct growth of large-area vertically stacked two-dimensional (2D) van der Waal (vdW) materials is a prerequisite for their high-end applications in integrated electronics, optoelectronics and photovoltaics. Currently, centimetre- to even metre-scale monolayers of single-crystal graphene (MLG) and hexagonal boron nitride (*h*-BN) have been achieved by epitaxial growth on various single-crystalline substrates. However, in principle, this success in monolayer epitaxy seems extremely difficult to be replicated to bi- or few-layer growth, as the full coverage of the first layer was believed to terminate the reactivity of those adopting catalytic metal surfaces. Here, we report an exceptional layer-by-layer chemical vapour deposition (CVD) growth of large size bi-layer graphene single-crystals, enabled by self-transmitting catalytic activity from platinum (Pt) surfaces to the outermost graphene layers. *In-situ* growth and real-time surveillance experiments, under well-controlled environments, unambiguously verify that the growth does follow the layer-by-layer mode on open surfaces of MLG/Pt(111). First-principles calculations indicate that the transmittal of catalytic activity is allowed by an appreciable electronic hybridisation between graphene overlayers and Pt surfaces, enabling catalytic dissociation of hydrocarbons and subsequently direct graphitisation of their radicals on the outermost $sp^2$ carbon surface. This self-transmitting catalytic activity is also proven to be robust for tube-furnace CVD in fabricating single-crystalline graphene bi-, tri- and tetra-layers, as well as *h*-BN few-layers. Our findings offer an exceptional strategy for potential controllable, layer-by-layer and wafer-scale growth of vertically stacked few-layered 2D single crystals.**

Controlled growth of AB-stacked bilayer (BLG) or few-layer graphene (FLG) is of fundamental interest because of their predicted physical properties toward potential applications in electronics and photonics [1-4]. Currently, metal-catalysed chemical vapour deposition (CVD) is one of the most effective methods to fabricate BLG and FLG on various substrates, including Ni[5,6], Cu[7-10], Ru[11-13], Ir[14,15], Pt[16], Cu-Ni [17,18], Pt-Si[19] and Cu-Si[20] alloys. In principle, additional graphene layer could be formed either above [21] or inserted below [22,23] an already grown graphene layer, corresponding to two types of the second layer, namely ad-layer (above) and intercalating layer (insert below), respectively. Hydrocarbons in the flowing gas during CVD growth could intercalate into the MLG-metal-catalyst interface through wrinkles, domain edges and grain boundaries and/or defects of MLG and then be dissociated into radicals at the interface[24]. Surface segregation of metal dissolved carbon onto the MLG- metal-catalyst interface provides an alternative carbon source but is limited by the solubility of carbon within the metal catalyst [18]. Regardless of the carbon source, growth at the MLG-metal-catalyst interface suffers from appreciable metal surface adhesion and strong metal-graphene interface confinement, which often leads to discontinuous carbon feeding, point or line defects, and/or unexpected coalescence boundaries of the grown intercalating graphene layers[19]. Recently, direct nucleation of a bilayer and subsequent formation of a large-area BLG on a metal surface was reported [25]. However, it is extremely technically challenging to control the number of graphene layers precisely and to reach a seamless-coalescence growth condition[25].

Layer-by-layer, namely Frank-van der Merwe (FM) is, in principle, an ideal and one of the most feasible strategies to prepare large-size BLG or FLG single-crystals, which was, during the preparation of this manuscript, realised using Cu foils in a bottom feeding and interface growing manner [26]. Although it is easier to monitor and control, the FM growth of graphene ad-layers on open surfaces (Fig. 1a) is, however, extremely challenging utilising the gas-phase feeding and open-surface growing manner because of a self-limiting growth effect that overwhelmingly favors monolayer formation [27]. Here, we demonstrate an FM growth of large-size single-crystalline graphene ad-layers (BLG and FLG), termed ``ad-layer FM growth'', directly on fully MLG covered Pt(111) by CVD methods. The choice of the Pt(111) surface enables vertically transmitting its catalytic activity onto the outermost graphene surface through graphene-Pt electronic hybridisation.

With the transmitted catalytic activity, the outermost carbon layer dissociates hydrocarbons on it, giving rise to formation of FLG following the FM growth mode, seamless coalescence of adlayers and healing of defective individual layers into continuous single-crystal over a large-scale.

Catalytic performance of metals to dissociate hydrocarbons was usually believed to abruptly drop after being covered with carbon layers [28,29]. This claim is, most likely, true for widely used Cu in growing graphene and *h*-BN monolayers [30,31]. However, other transition metals that are conventionally used for metal catalysts in molecular reactions could offer much stronger hybridisation with the MLG. It is yet to be explored whether such hybridisation enables transmitting the catalytic activity of metal *d*-electrons onto the carbon surface through MLG. If the answer is positive, the MLG thus inherits the metal catalytic activity, being capable of dissociating coming hydrocarbons. While Ru, Rh or other active transition metals provide stronger hybridisation with MLG than Cu, the reinforced MLG-metal interaction results in significant surface Moiré corrugation of the MLG on those surfaces, which suppresses the diffusion of radicals and thus hinders the growth of graphene adlayers[32,33]. In addition, the relatively high carbon solubility of those metals usually leads to carbon segregation at the MLG-metal interfaces, which may compete with the growth process on the outermost surface, leading to the growth behaviour being less controllable. The MLG-Pt(111) interface seems unique among those MLG-metal-catalyst interfaces because of nearly flat MLG on Pt(111)[34], relatively low carbon solubility in Pt and an appreciably strong C-Pt hybridisation [32,35].

We first theoretically consider the MLG-Pt(111) interface using a MLG-(9×9)/Pt(111)-(8×8)-R0° supercell (Extended Data Fig. 1a), which was found to be the most experimentally common configuration for this interface [36,37]. A projected and unfolded electronic band structure illustrates the MLG-Pt hybridisation details (Fig. 1b). While it clearly shows a blue cross representing the Dirac cone sitting at ~0.35 eV above the Fermi level ($E_F$), substantially flatter Pt *d* bands (light blue lines) that appreciably hybridise with the carbon $p_z$ component were identified both near $E_F$ and around roughly -1 eV below $E_F$. Visualised wavefunction norm (Fig. 1c) of a hybridised state near $E_F$ further verifies the hybridisation; this shows that electronic states of the Pt-supporting MLG surface exhibit *d*- characteristics near $E_F$. We also plotted the density of states revealed by our DFT calculations and an experimental scanning tunneling spectroscopic (STS) spectrum of MLG-(9×9)/Pt(111)-(8×8)-R0° in Fig. 1d. They share several features, e.g. comparable positions of the Dirac point at 0.30 eV (STS) and 0.35 eV (DFT), denoted using the blue and red triangles, respectively and nearly superposed hybridised states around $E_F$ (the dark gray arrows in Fig. 1b and 1d). These consistent results indicate that the $d$ characteristics arising from electrons of Pt could be found for electrons distributed on the graphene layer through this Gr-Pt electronic hybridisation. The existence of $d$ characteristics, which usually lowers dissociation barriers of hydrocarbons, over the outermost graphene layer implies the potential catalytic activity of MLG/Pt(111) towards dissociation of coming hydrocarbons onto MLG.

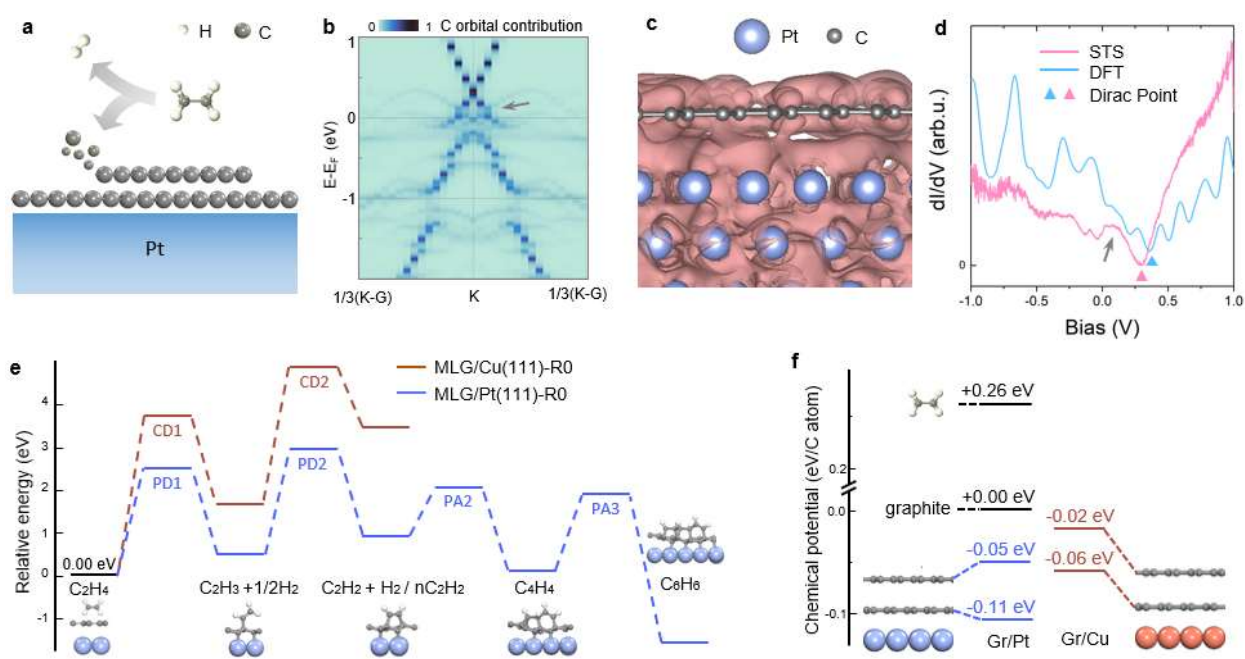

**Figure 1 | Theoretical evidence to support the FM growth-mode of BLG on Pt(111). a**, Sketch of an effective FM CVD growth mode of BLG on Pt. **b,** Projected band structure of MLG/Pt(111)-R0°. Those blue blocks stand for the $p_z$ orbital population on all C atoms and the depth of the colour represents the weight of the $p_z$ component. **c**, Visualised wavefunction norm square of a representative hybridised state of MLG/Pt(111)-R0° near the Fermi level. The isosurface value used in the plot is $1×10^{-5}$ $e$/Bohr$^3$. **d,** Experimental STS spectrum (red) and a corresponding theoretical plot (blue) of the density of states of MLG/Pt(111)-R0°. The Dirac point (DP) is marked by the red (expt.) and blue (theor.) triangles, respectively, while C-Pt hybridised states near the Fermi level are indicated using the grey arrow. **e**, Most likely C$_2$H$_4$ dissociation and nucleation pathways, including reaction barrier heights, intermediate structures and their energies, for the on-surface nucleation on MLG on Pt(111) (blue) and Cu(111) (red). Here, letters ``P'' and ``D'' represent ``Pt'' and

``Dissociation'', respectively and PD1(2) thus denotes the first (second) transition state for $C_2H_4$ dissociation on MLG/Pt(111), while, given ``C'' for ``Cu'', CD1(2) stands for that on MLG/Cu(111). Given ``A'' for the first letter of term ``Aggregation'', PA2(3) represents the transition state of two (three) $C_2H_2$ radicals aggregating together on MLG/Pt(111). **f**, Chemical potentials of each C atom in ethylene, graphite, M(B)LG/Pt(111)-R0° and M(B)LG/Cu(111), respectively.

Figure 1e shows the reaction pathways and their associated reaction barriers of graphene nucleation on the surface of MLG/Pt(111), which indicates the $C_nH_n$ species is preferred in the nucleation and expansion processes. A $C_2H_4$ molecule, the carbon source used in our experiments, releases 0.21 eV upon physisorption on the MLG. This physisorption state serves as the zero-energy reference. The pathway in blue (the Pt series) shows that the adsorbed $C_2H_4$ molecule surmounts a 2.50 eV barrier [PD1, the $1^{st}$ dissociation barrier of $C_2H_4$ on Pt(111)] to break a C-H bond, forming a $C_2H_3$ radical. That is followed by a 2.93 eV barrier (PD2) to form an adsorbed $C_2H_2$ radical. The barrier for breaking two C-H bonds of the same C atom (Extended Data Fig.1) is over 4.06 eV, making this pathway unlikely. Compared with the corresponding barriers on the Cu surface, e.g. 3.69 (CD1) and 4.85 eV (CD2), the stronger C-Pt hybridisation substantially reduces the barriers and activates the catalytic ability of MLG. Polymerisation of $C_2H_2$ radicals into $C_nH_n$ chains is much more preferred with barriers of roughly or less than 2 eV (PA2 and PA3). At the same time, continuous breaking of the C-H bonds needs to surmount barriers twice PA2 or PA3, i.e. 4.06 and 5.91 eV for forming $C_2H$ and $C_2$, respectively (Extended Data Fig. 1). The preferred $C_nH_n$ species is thus the primary form of radicals moving on the MLG, consistent with the previous finding for MLG growth on Cu surfaces[21].

The MLG/Pt(111) surface does show catalytic activity to dehydrogenate hydrogen carbons. Given the overall barrier of 2.93 eV for the most likely pathways on MLG/Pt(111), dehydrogenation and nucleation processes are predicted to occur over 850°C, highly consistent with our experimentally observed initialisation temperature of 900 °C. However, the same processes on Cu(111) requires a temperature of near 1400 °C to initialise, which largely exceeds the melting point of Cu(111). Indeed, the formation of an adlayer is barely observed even on liquid copper surfaces.[38] Chemical potential analysis also supports the difficulty of growing BLG on Cu(111) surface, as shown in Fig. 1f. The chemical potential of the $2^{nd}$ layer Gr on Cu(111) is only -0.02 eV, very close to that of graphite, which is taken as a zero reference for the chemical potential. In comparison, the potential level of BLG lowers to -0.05 eV on Pt(111). This appreciably lowered potential level indicates reinforced interlayer interaction of BLG with the Pt surface. Our simulations show that the carbon-platinum hybridisation, the reduced dissociation barrier of hydrocarbons and the lowered chemical potential are not limited to the MLG-(9×9)/Pt(111)-(8×8)-R0° interface. We have observed that these effects are maintained at another MLG-Pt interface, namely MLG-(3×3)/Pt(111)-($\sqrt{7}\times\sqrt{7}$)-R19° (Extended Data Fig. 2), which could be found under some experimental conditions as well.[34] Additionally, our DFT calculation shows that such strong carbon-platinum hybridisation still exists in BLG/Pt(111) and 3L-Graphene/Pt(111) using the FLG-(3×3)/Pt(111)-($\sqrt{7}\times\sqrt{7}$)-R19° models which are computationally less demanding (Extended Data Fig. 3). Thus, all these results support the inheritance of the catalytic activity arising from Pt onto the outermost-layer of FLG/Pt(111), suggesting a likely route to the ad-layer FM growth of FLG, as we elucidate later.

Next, the theoretically predicted ad-layer FM growth mode of BLG (Fig. 2a) was verified by our near atmosphere pressure (NAP) experiments as follows. A MLG sample was first prepared using a $C_2H_4$ carbon source on a Pt(111) surface in a modified environmental scanning electron microscope (ESEM)[39] (Fig. 2a and 2b). By tuning the $C_2H_4/H_2$ ratio under ESEM surveillance, we explicitly observed that MLG domains, which are hexagonal in shape and are oriented parallel to each other, nucleate at moment $t_0$ (Fig. 2b), expand and eventually stitch to each other at $t_0$+8s (Fig. 2c), demonstrating a highly aligned growth mode of the first carbon layer (MLG). At the same moment ($t_0$+8s), an adlayer domain (BLG) begins to form (Fig. 2c) on the as-grown MLG surface and expands (Fig. 2d) in the same orientation until, at least, $t_0$+31s with a smaller growth rate (Fig. 2e). The slower growth rate of the second layer evidences a larger dissociation barrier of $C_2H_4$ on MLG/Pt(111) in comparison with that on bare Pt(111), qualitatively consistent with the larger barrier of 2.93 eV and the smaller barrier of 1.66 eV for both surfaces, respectively. In addition, the almost constant growth rates of MLG and BLG (Fig. 2e) and the unobserved fractal structural domains imply the expansion stays in the attachment-limited growth mode[39] and the dehydrogenated radicals are sufficient for both MLG and BLG growth. A continuous BLG film eventually forms over a prolonged period of growth time.

It is of crucial importance in distinguishing this work from other growth strategies that whether the growth fronts are on top of the MLG/Pt(111). It was experimentally verified in two ways. An *in-situ* etching experiment was conducted by slightly lifting the chemical potential level of BLG, i.e. increasing the H$_2$ partial pressure, in which the hexagonal ad-layer (second layer) flake shown in Fig. 2f begins to shrink at time $t_{0'}$ and develops to a smaller and less regular flake at time $t_{0'}+43s$ (Fig. 2g), while the first layer maintains intact. This observation proves that nucleation of the ad-layer graphene occurs on top of MLG/Pt(111) (the first layer) and is thus exposed to the etching atmosphere (Fig. 2f-g) [40]. This *in-situ* observation thus compellingly supports the ad-layer FM growth of BLG on Pt(111). Additionally, a cross-sectional view of such growth and etching process is provided using *in-situ* transmission electron microscopy (TEM) (see Methods for details). Figure 2h depicts three representative snapshots and their schematic illustrations from a real-time visualisation (Supplementary Movie 1), which clearly shows graphene adlayers nucleate and expand on the top of the fully covered first graphene layer. An etching process, as expected, occurs primarily on the outermost layer under a pure hydrogen atmosphere, as shown in Extended Data Fig. 4 and Supplementary Movie 2.

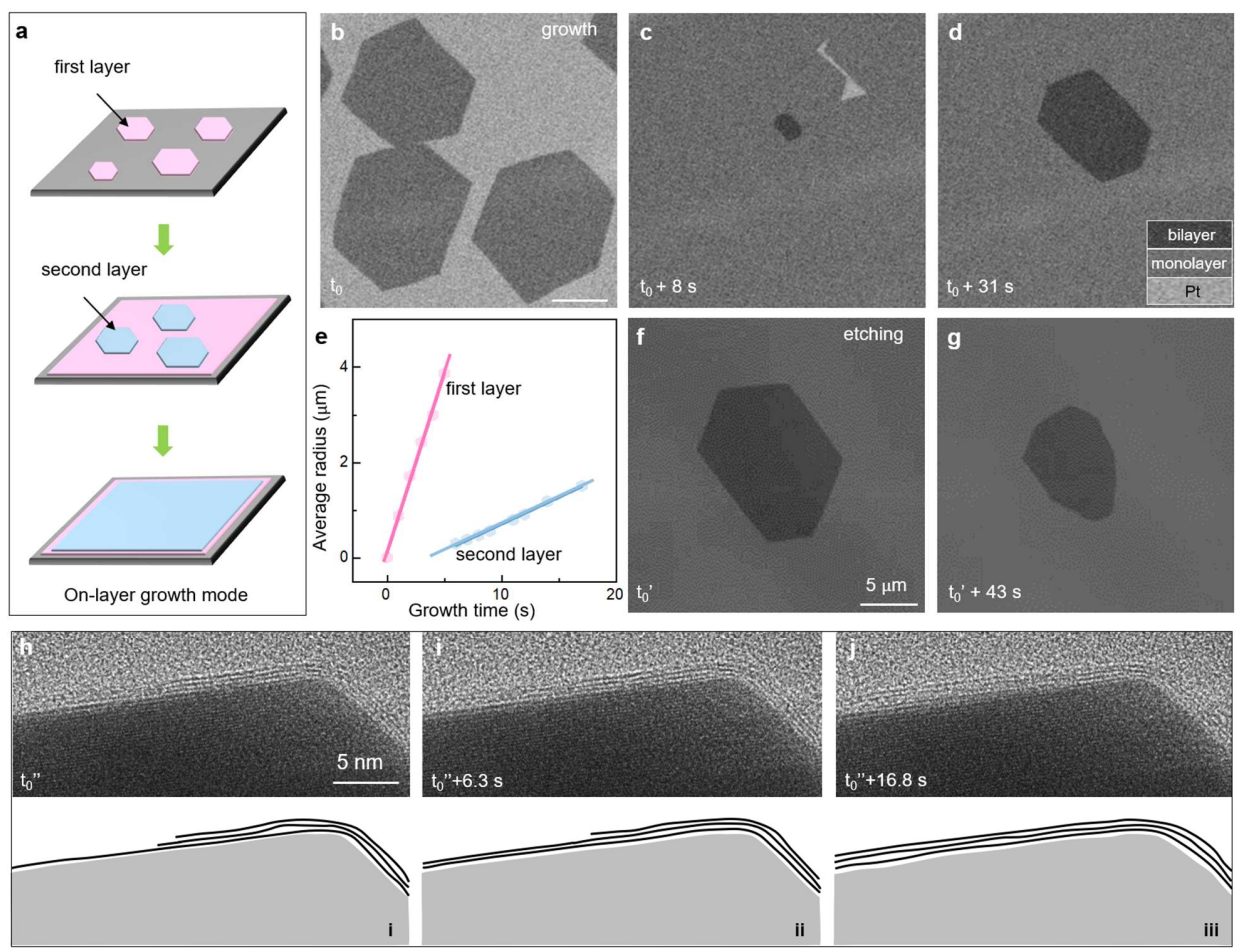

**Figure 2 | In-situ observation of BLG growth on Pt in a layer-by-layer growth mode. a,** Schematics of the bilayer graphene growth in a layer-by-layer growth mode. **b-d**, ESEM images of MLG domains (a), nucleation (b) and lateral growth (c) of the second layer (BLG) at times $t_0$, $t_0 + 8$ and $t_0 + 31$s, respectively. The MLG domains are well aligned, and the BLG domain shares the same orientation with the MLG ones. **e**, Averaged radii of the MLG and BLG domains as a function of growth time. The domains radii of both layers show a linear dependence on the growth time, with rates of 0.75 and 0.11 μm/s, respectively. **f-g**, Etching of the BLG domain. After hydrogen etching gets started, the BLG domain begins to shrink while the MLG film remains intact, demonstrating the second layer is grown on the top of the first layer. **h-j**, Cross-sectional *in-situ* TEM images (upper panels) and their schematics (lower panels) show the FM growth of graphene ad-layers on the surface of a Pt particle at different growth times.

The nucleation and subsequent growth of the second layer are inevitably accompanied by nucleation of third- or thicker-layer islands, because of the comparable (within 0.05 eV) chemical potential levels of those ad-layer islands. Figure 3a-3d shows a formation process of MLG where thicker-layer islands were nucleated, which are, however, undesired. Nevertheless, they are removable by precise keeping the relative chemical potential ($\mu$) of graphene layers and waiting for the sample to approach its dynamic equilibrium, during which defective regions are spontaneously replaced by perfectly oriented lattices driven by the thermal stability of the latter, as shown in Fig. 3e-3h. We further term this process ``vertical healing (VH)''.

In the meantime, another healing type, i.e. ``lateral healing (LH)'', occurs within layers, which could both remove point defects and eliminate line defects formed in domain coalescence, as examined by etching the samples under $H_2$ atmosphere. Figure 3i shows a fast-grown MLG sample where the MLG nucleates from hundreds of domains (Fig. 3b) and expands into monolayer before other layers start to nucleate. Massive and irregular holes (Fig. 3j) are observable under direct etching of the sample, and it eventually fragments (Fig. 3k and 3l), suggesting its quality is inferior. However, the MLG shown in Fig. 3h was treated with both VH and LH processes, which was replotted in Fig. 3m. There are only three pin-holes found in this sample after 10 s direct etching. They subsequently merge into two larger holes and eventually into a single hole after 2640 s etching, suggesting large quality improvement.

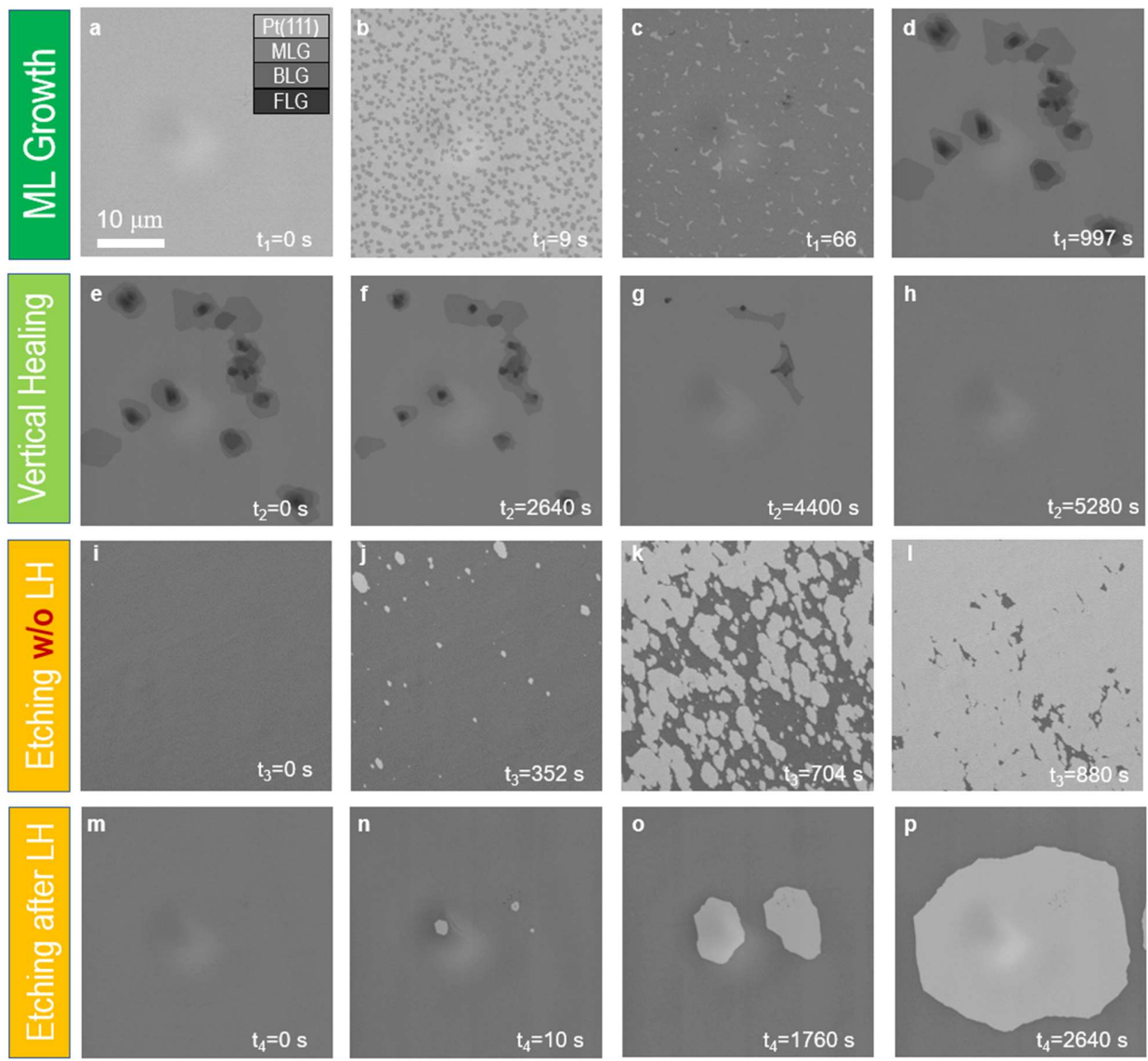

**Figure 3 | Real-time ESEM images showing vertical and lateral healing processes of as-grown MLG on Pt(111) at NAP. a-d**, a process showing MLG growth from a bare Pt(111) surface (a), a formation of hundreds of nuclei (**b**), nearly fully covered MLG (**c**), and eventually to a fully covered MLG with FLG islands (**d**). ML stands for monolayer. The growth is governed by kinetics and off-equilibrium. Those FLG islands could be vertically healed under the same condition when the sample nearly reaches equilibrium, as illustrated in a typical process (**e-h**). Panel **e** shows an image of an as-grown MLG with multi-layer islands, shrinking of those FLG islands (**f** and **g**) and their final disappearance(**h**). During this period MLG is also laterally healed. Panels **i** to **l** show an etching process of a fast-grown MLG , without both VH and LH. Panel **i** depicts the MLG sample while panels **j-l** show the subsequent etching process. Panel **h** was replotted in panel **m**, which represents a thermodynamically stable structure after both the VH and

LH processes, while the subsequent etching process, with the same etching condition, is shown in panel **n-p**.

In terms of BLG samples, Extended Data Fig. 5, remarkably, shows *in-situ* observations of etching before and after healing. The line etched between two hexagonal pin-holes is indicative of line defect presence in the sample before LH (Extended Data Fig. 5a and 5b). After LH, the absence of the etched line defects (within an inspected area of 500 $\mu m^2$) indicates the sample is, most likely, free of line defects (Extended Data Fig. 5c-5e). Additionally, aligned shapes of those hexagonal pin-holes suggest that the BLG film is a single-crystal.[41] Supplementary Movie 3 illustrates all those processes, i.e. nucleation, expansion, etching and healing, of FLG at NAP.

To further validate the above discussion and generalise the layer-by-layer mechanism from NAP to UHV (Ultra-High Vacuum), we performed additional *in-situ* SEM observation, reproducing the BLG growth, healing and etching processes in a *one-round continuous* experiment utilising the temperature variation strategy in UHV. Extended Data Fig. 6 illustrates the growth and etching processes of the first and second layers using totally 12 snapshots from a complete movie (Supplementary Movie 4) recording the experimental procedure, in which the third healing type, i.e. ``regrowth healing (RH)'' is clearly depicted for MLG. The RH is an accelerated process of VH and/or LH that the defective region is locally etched, promoting subsequent regrowth at near equilibrium. The local etching could be realised by either increasing the $H_2$ partial pressure at NAP or heating the substrate for reducing the concertation of $C_nH_n$ radicals on the surface in UHV. The latter strategy utilises the adsorption-desorption balance of $C_2H_4$ molecules and hydrocarbon radicals, which was employed in the process shown in Supplementary Movie 4.

We magnified those parameters obtained from our *in-situ* observations to fit the size of our tube furnace systems for further verifying the generality of our strategy; this allows the ad-layer FM growth of BLG (FLG) to be accessed by the conventional CVD method (see Methods for more details). Figure 4a and Extended Data Fig. 7 show SEM and optical images of as-grown BLG domains, respectively, which are in a hexagonal shape with aligned edges. These domains coalesce and eventually form a uniformly continuous BLG film after a prolonged growth time (Fig. 4b).

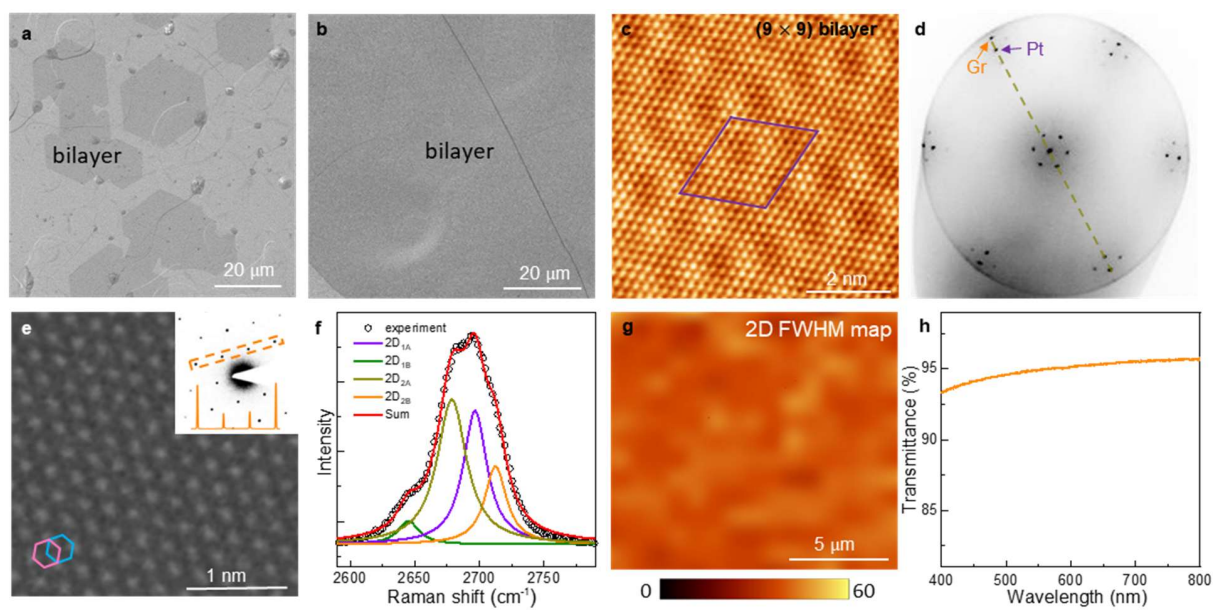

**Figure 4 | Tube-furnace CVD growth of bilayer graphene on Pt(111). a-b**, SEM images of well-aligned second layer graphene domains (a) and continuous BLG film (b). **c-e**, STM image (**c**), LEED pattern (**d**) and STEM image (**e**) of the as-prepared BLG sample. Inset in (e): SAED pattern and one set of intensity profiles. **f**, Lorentzian fit of the 2D peaks for the BLG sample, indicating the sample contains AB-stacked BLG. **g**, Raman FWHM map of the 2D peak, confirming the sample is comprised of AB-stacked BLG. **h**, Transmittance spectrum of an as-grown BLG film transferred onto a quartz substrate, with a transmittance of ~ 94.9% at 550 nm.

Atomic-resolution scanning tunnelling microscopy (STM) images of both MLG/Pt(111) and BLG/Pt(111) show the 9×9-R0° Moiré pattern (Fig. 4c and Extended Data Fig. 8). The alignment of BLG and Pt(111) lattices is supported by the low energy electron diffraction (LEED) pattern shown in Fig. 4d. Scanning transmission electron microscopy (STEM) imaging and selected area electron diffraction (SAED) patterning (Fig. 4e and Extended Data Fig. 9) further prove the desired AB-stacked[42], single-crystalline and defect-free BLG. The high quality and uniformity of BLG were also examined using optical measurements. We used a previously reported electrochemical bubbling method to transfer our tube-furnace grown BLG onto a SiO$_2$/Si substrate.[43] Figure 4f shows the characteristic 2D peak of a representative Raman spectrum of the transferred BLG sample. This peak is asymmetric and is fitted using four Lorentzian peaks at different frequencies. A full-width-at-half-maximum (FWHM) mapping of the 2D peak is displayed in Fig. 4g, offering a uniform FWHM of about 50 cm$^{-1}$ across tens of microns, again, verifying the AB-stacking of our BLG sample.

In addition, an optical transmittance of ~ 95.2% at 500 nm was achieved in a BLG film being transferred onto a quartz substrate; conforming its bilayer nature[44] (Fig. 4h).

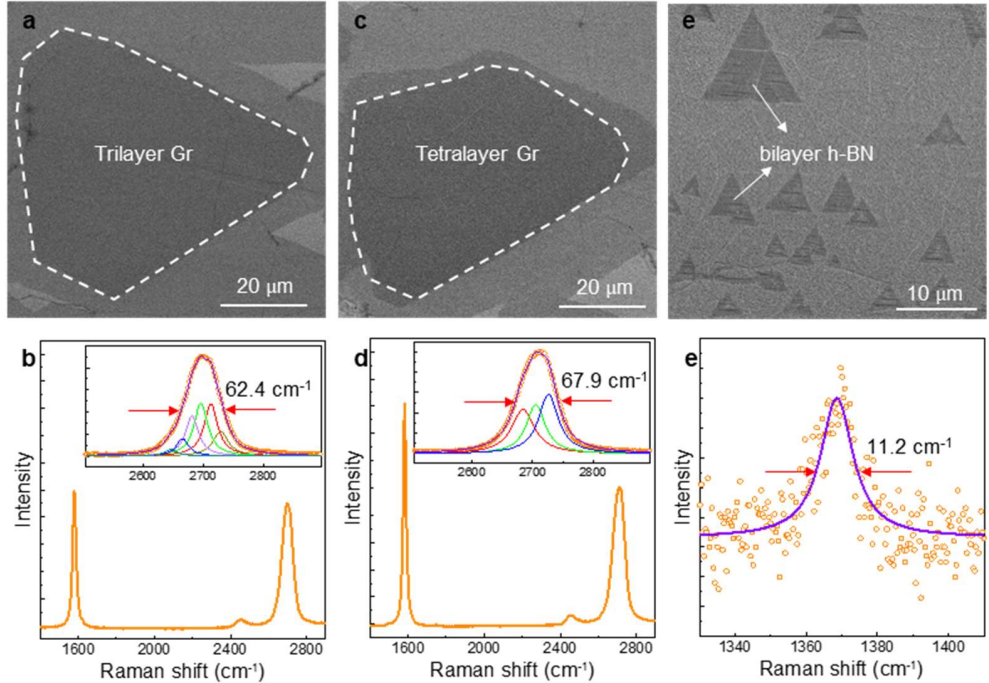

**Figure 5 | Tube-furnace CVD growth of 3L-, 4L-graphene and bilayer *h*-BN. a-f,** SEM images (**a,c,e**) and corresponding Raman spectra (**b,d,f**) of 3L-, 4L-graphene and bilayer *h*-BN. Inset in (**b,d**): Lorentzian fit of the 2D peaks for AB-stacked 3L- and 4L-graphene, respectively.

All those experimental results represented above demonstrate that the catalytic activities of Pt could indeed penetrate through MLG and enable the FM growth of BLG as our theory predicted. Meanwhile, our calculations also identified the catalytic activity on the outermost carbon layer of BLG/Pt(111) and 3L-Graphene/Pt(111), which was experimentally realised by increasing the $C_2H_4$ partial pressure. As shown in Fig. 5a-5d, 3L- and 4L-graphene were successfully prepared. Raman characterisations verified the numbers of layers and the AB stacking order. Bilayer *h*-BN thin-films can also be grown in the ad-layer FM mode on Pt(111), suggesting the transmitting catalytic activity on monolayer *h*-BN inheriting from Pt(111) (Fig. 5e). Figure 5f highlights the FWHM, i.e. 11.2 cm$^{-1}$, of a specific Raman peak of the *h*-BN bilayer, which is even smaller than those usually measured in few-layer CVD *h*-BN (14-20 cm$^{-1}$)[45,46], suggesting, most likely, higher quality of the ad-layer FM grown bilayer *h*-BN than other CVD prepared samples.

In summary, we theoretically predicted the ad-layer FM-mode growth of AB-stacked single-crystal BLG on Pt(111), as well as FLG and bilayer *h*-BN, which were experimentally verified using *in-situ* CVD under NAP and UHV, respectively, and tube-furnace CVD. Real-time observation of evolving adlayers under relevant CVD conditions provides the missing clue for unravelling the transmitted activity of metal catalysts onto the outermost graphene surface through coupling $sp^2$ carbon, as predicted by our DFT calculations. An interlayer resonance bonding state comprised of C *p* and Pt *d* orbitals of FLG, in principle, enables a long-distance delivery of C-Pt interfacial state to the outermost graphene surface. In other words, it suggests that the catalytic activity of Pt(111) could persist but with reduced strength on the FLG/Pt(111) surface. The uniqueness of the present case lies in the simultaneous imaging of both catalyst (*n*-layer graphene) and products (*n*+1-layer graphene). This particular feature allows direct verification of a very recently proposed viewpoint of ``chainmail'' catalysts which claims delivery of catalytic activity of encapsulating metals to the surface of their surrounding $sp^2$ carbons [47]. Moreover, this exceptional growth strategy is, in principle, expected to have high promise in fabricating other 2D nanostructures on other metals, especially with emphases of high controllability, the capability of healing, direct surveillance, line-defect free and single-crystalline.

## Methods

**Density functional theory calculation.** Our first-principles density functional theory (DFT) calculations were carried out using the projector-augmented wave (PAW) potential approach[48], the revised Perdew-Burke-Ernzerhof (RPBE)[49] flavour of the generalised-gradient-approximation (GGA)[50] for the exchange-correlation potential and a plane-wave basis set as implemented in the Vienna Ab-initio Simulation Package (VASP)[51]. Van der Waals interactions were considered using the DFT-D3 method[52] and double-checked with the optB86b-vdW functional [50,53] for structure relaxations and electronic structure calculations. Kinetic energy cut off for the plane-wave basis set is set to 400eV. A lattice constant of 2.7876 Å for Pt was used and was kept fixed during structural relaxations. A slab model consisting of four layers of Pt atoms and a 20 Å vacuum layer was employed to model the Pt(111) surface. One or more graphene layers were placed on the Pt slab with rotational angles of 0 and 19 degrees, forming a 9×9-Gr/8×8-Pt(111) and a 3×3-Gr/$\sqrt{7}\times\sqrt{7}$-Pt supercell, respectively. A Gamma point and a k-mesh of 5×5×1 were used for geometry optimisation. In electronic structure calculations, denser k-meshes of 3×3×1 and 9×9×1 were adopted. The shape and volume of the supercells were kept fixed during relaxation, and all atoms except the two bottom layers of Pt were allowed to fully relax until the residual force on each atom is less than 0.01 eV/Å. Chemical potentials were derived by comparing the formation enthalpy of each species while those of graphite, hydrogen gas and platinum surface were set to reference zero. We used the KPROJ program[54] based on a k-projection method to unfold those highly folded supercell bands into unit cell ones. A k-point route from Γ(0,0,0) to K(3,3,0) was used for the electronic band unfolding. Reaction barriers were revealed using the nudged elastic band theory calculations [55] with a criterion force of 0.02 eV/Å

**In-situ CVD growth, healing and etching of FLG on Pt(111) at NAP under ESEM surveillance.** In-situ CVD growth experiments were performed inside the chamber of a modified commercial ESEM (Thermofisher Quattro). The vacuum system of the ESEM was modified and upgraded with oil-free roughing pumps. The instrument is equipped with a homemade laser heating stage, a gas supply unit (mass flow controllers from Bronkhorst), and a mass spectrometer (Pfeiffer OmniStar) for online analysis of the chamber atmosphere. Owing to the use of rubber O-rings for the sealing and the fact that the chamber cannot be baked out, the base pressure of the instrument is around $4 \times 10^{-6}$ Pa. Under CVD growth conditions, the pressure is eight orders of magnitude higher than the base pressure and constitutes mostly of $H_2$ (99.9995% purity) and $C_2H_4$ (99.95% purity). No influence of the electron beam on the growth and etching process could be observed. The imaged regions and their respective surroundings showed similar behaviour, as evidenced by changing the magnification or by moving the sample under the beam. Furthermore, no electron beam induced contamination was observed at elevated temperatures.

**Tube-furnace CVD growth of graphene on Pt(111).** A single-crystal Pt(111) substrate was specifically pretreated to remove its dissolved carbon atoms before the FM growth of graphene adlayers. Then the Pt(111) substrate was first annealed at 1000 °C for 40 min under reducing atmosphere. Subsequent growth of BLG was conducted in two steps: $CH_4$ was first fed into the system to grow the first graphene layer. With the full coverage of the first layer, $CH_4$ was cut off and $C_2H_4$ was introduced to activate the growth of the second graphene layer. The whole growth process was conducted at atmospheric pressure.

*In-situ* **TEM characterisation.** Pt nanoparticles (Adamas-beta) were dispersed in pure ethanol. The mixture was subsequently drop-casted on a 2mm × 2mm microelectromechanical system (MEMS) chip with heating membrane provided by Protochips Inc (Morrisville, NC, USA) followed by drying in air. The as-prepared chip is paired with another 4.5 mm × 6 mm chip with e-beam transparent $SiN_x$ window to compose the in situ environmental reaction cell. An electrical current is applied through the gold contacts to the SiC heating membrane of the bottom chip, which contains six holes of 8 μm diameter layered with a thin $SiN_x$ film to facilitate sample observation in TEM[56]. Due to the small size of the heating area, a homogeneous temperature distribution is obtained within the analysed areas. Thereby, it is feasible of heating the sample up to 1000 °C with a temperature rate up to 10 °C/s. The chips were then loaded with gaskets onto an in situ TEM holder provided by Protochips to ensure hermetic sealing from the TEM vacuum. A 5μm spacer bridges the top and bottom chips for the cell. A flowing gas was applied to the cell from an external gas feeding system. An integrated residual gas analyser (RGA) was used to verify the gas mixture applied to the cell[57]. An aberration-corrected JEM GrandARM 300F transmission electron microscope (operated at 300 kV) was used to carry out in situ graphene growth experiments. The Pt nanoparticles were treated at 700 °C in 10%$H_2$/90%Ar gas (1 bar, 0.1 sccm) for one hour. In situ observations of graphene growth under CVD conditions were made at 950°C with a gas mixture of 0.5%$C_2H_4$/9.95%$H_2$/89.55%Ar gas (1 bar, 0.1 sccm). Observations of graphene etching under 0.5%$C_2H_4$/9.95%$H_2$/89.55%Ar gas (1 bar, 0.1 sccm) were carried out at 900°C. Sequential bright field TEM images were recorded in situ by Gatan OneView IS Camera at acquisition rates of 4-25 frames per second at 2K × 2K frame resolution.

**Ex-situ characterisations.**

**SEM and STEM:** SEM images were obtained using Thermal Fisher Quattro S system. STEM and TEM experiments were performed in FEI Titan Themis G2 300 operated at 80 kV.

**STM/STS:** The BLG/Pt(111) sample was heated at ~300 °C for 3 hrs after loading into the UHV chamber (base pressure $2 \times 10^{-10}$ mbar) for degassing, and then the sample was transferred into the STM head (Createc) to perform the measurements. STM imaging was performed using mechanically cut PtIr tips at the liquid helium temperature (4.2 K). The d$I$/d$V$ tunnelling spectra were acquired using lock-in detection by applying an AC modulation of 20 mV (r.m.s.) at 973 Hz to the bias voltage.

**Raman:** Raman spectra and mappings were collected with a homemade optical system using a 532 nm laser excitation. The transmittance was measured using a Varian Cary 5000 UV-Vis-NIR spectrophotometer.

**LEED:** low-energy electron diffraction was conducted in a SPECS low-energy electron microscopy system, which is connected to the Vacuum Interconnected Nanotech Workstation (NANO-X) of Suzhou Institute of Nano-Tech and Nano-Bionics, the Chinese Academy of Sciences, which is installed with a preparation chamber and the main chamber. LEED was performed to investigate surface structure after graphene growth. The incident electron beam energies are always kept at 50 eV during LEED measurements.

**Acknowledgements:**

This work was supported by the National Key R&D Program (Grant No. 2018YFE0202700) and Key R&D Program of Guangdong Province (Grant Nos. 2018B030327001, 2019B010931001 and 2020B010189001), the National Natural Science Foundation of China (11622437, 11888101,

11974422, 51991342, 52021006, 52025023, 61888102 and 61674171), the Strategic Priority Research Program of Chinese Academy of Sciences (XDB30000000, XDB33000000), Beijing Natural Science Foundation (JQ19004), the Pearl River Talent Recruitment Program of Guangdong Province (2019ZT08C321), China Postdoctoral Science Foundation (2020T130023). The research was supported by European Commission (H2020-Twinning project No. 810626–SINNCE) and Brno University of Technology (Specific research No. FSI-S-20-6485). Calculations were performed at the Physics Lab of High-Performance Computing of Renmin University of China, Shanghai Supercomputer Center. The authors are grateful for the technical support for Nano-X from Suzhou Institute of Nano-Tech and Nano-Bionics, Chinese Academy of Sciences (SINANO).

**Author contributions**

Z.-J.W., K.L and W.J. conceive of this project and supervised the research; Z.-J.W. carried out *in-situ* CVD growth, performed ESEM and TEM imaging and, together with Z.C., Q.Z., Z.L, M.W. and W.J., analysis the data; Z.Z., R.Q., E.W. and K.L. conducted tube-furnace CVD growth, STEM and optical imaging, as well as Raman measurements; L.Z., C.Y., J.Q. and W.J. carried out all theoretical calculations and analysis; A.J., P.B and M. K.. performed CVD growth and in-situ SEM imaging under UHV and analysed the data; Q.Z., T.Z., L.L, L.B., H.Y., Z.C., Y.W. and H.-J.G. performed STM and STS experiments; W.W. and Y.C. conducted LEED measurements. Z.-J.W., Z.Z., L.Z., K.L. and W.J. write the manuscript with inputs from the all authors.

# Layer-by-layer growth of bilayer graphene single-crystals enabled by self-transmitting catalytic activity

Zhihong Zhang[1,2,#], Linwei Zhou[1,#], Zhaoxi Chen[3,4,#], Antonín Jaroš[5,6], Miroslav Kolíbal[5,6], Petr Bábor[5,6], Quanzhen Zhang[7], Changlin Yan[1], Ruixi Qiao[2], Qing Zhang[3], Teng Zhang[7], Wei Wei[8], Yi Cui[8], Jingsi Qiao[7], Liwei Liu[7], Lihong Bao[9,11], Haitao Yang[9,11], Zhihai Cheng[1], Yeliang Wang[7], Enge Wang[2], Zhi Liu[3,4], Marc Willinger[10], Hong-Jun Gao[9,11], Kaihui Liu[2,*], Zhu-Jun Wang[4,3,10,*], and Wei Ji[1,*]

[1] Beijing key laboratory of optoelectronic functional materials and micro-nano devices, Department of Physics, Renmin University of China, 100872, Beijing, China

[2] State Key Laboratory for Mesoscopic Physics and Frontiers Science Center for Nano-optoelectronics, School of Physics, Peking University, Beijing 100871, China.

[3] Center for Transformative Science, ShanghaiTech University, Shanghai 201210, China

[4] School of Physical Science and Technology, ShanghaiTech University, Shanghai 201210, China

[5] CEITEC BUT, Brno University of Technology, Purkyňova 123, 612 00 Brno, Czech Republic

[6] Institute of Physical Engineering, Brno University of Technology, Technická 2, 616 69 Brno, Czech Republic

[7] School of Information and Electronics, MIIT Key Laboratory for Low-dimensional Quantum Structure and Devices, Beijing Institute of Technology, Beijing 100081, China

[8] Vacuum Interconnected Nanotech Workstation, Suzhou Institute of Nano-Tech and Nano-Bionics, Chinese Academy of Sciences, Suzhou 215123, China

[9] Beijing National Laboratory for Condensed Matter Physics and Institute of Physics, Chinese Academy of Sciences, 100190, Beijing, China

[10] Scientific Center for Optical and Electron Microscopy, ETH Zurich, Otto-Stern-Weg 3, 8093 Zurich, Switzerland

[11] School of Physical Sciences, University of Chinese Academy of Sciences, Beijing 100039, China

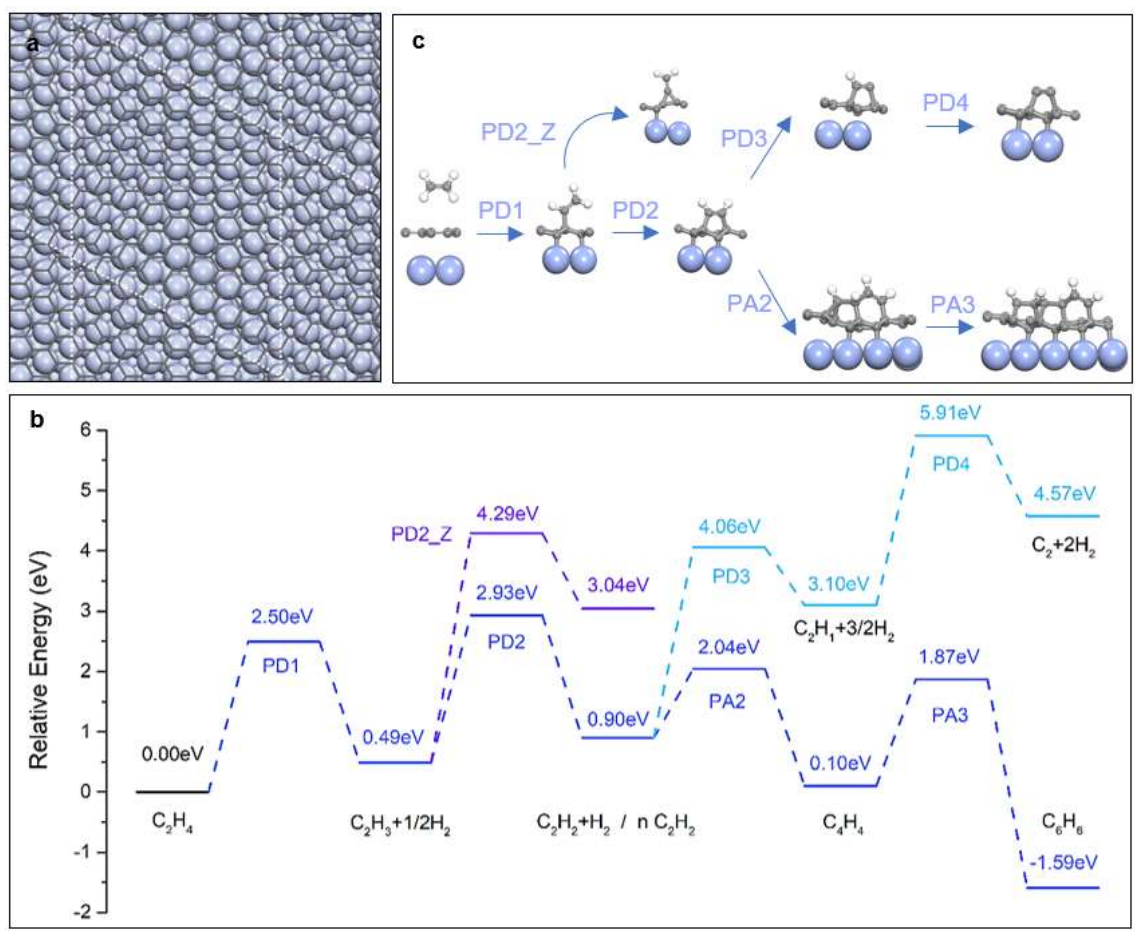

**Extended Data Fig. 1 | Reaction pathways for C$_2$H$_4$ dissociation and graphene nucleation on MLG-(9x9)/Pt(111)-(8x8)-R0°. a**, Atomic model of the supercell. **b**, three potential pathways and their corresponding barrier heights for the C$_2$H$_4$ dissociation reaction and subsequent radical nucleation. In each reaction step, relative energies of the initial- (IS), transition- (TS) and final-states (FS) were plotted and the energy level of the IS of a certain step is aligned to that of the FS in the last step. Atomic models of the reactant, product and their intermediates were shown in **c**, State PDn stands for the transition state of the nth H dissociation on MLG/Pt(111) and notation PAn represents the step of n radical aggregating reaction on the surface. Suffix Z denotes the dissociation of two H atoms at the same C atom site.

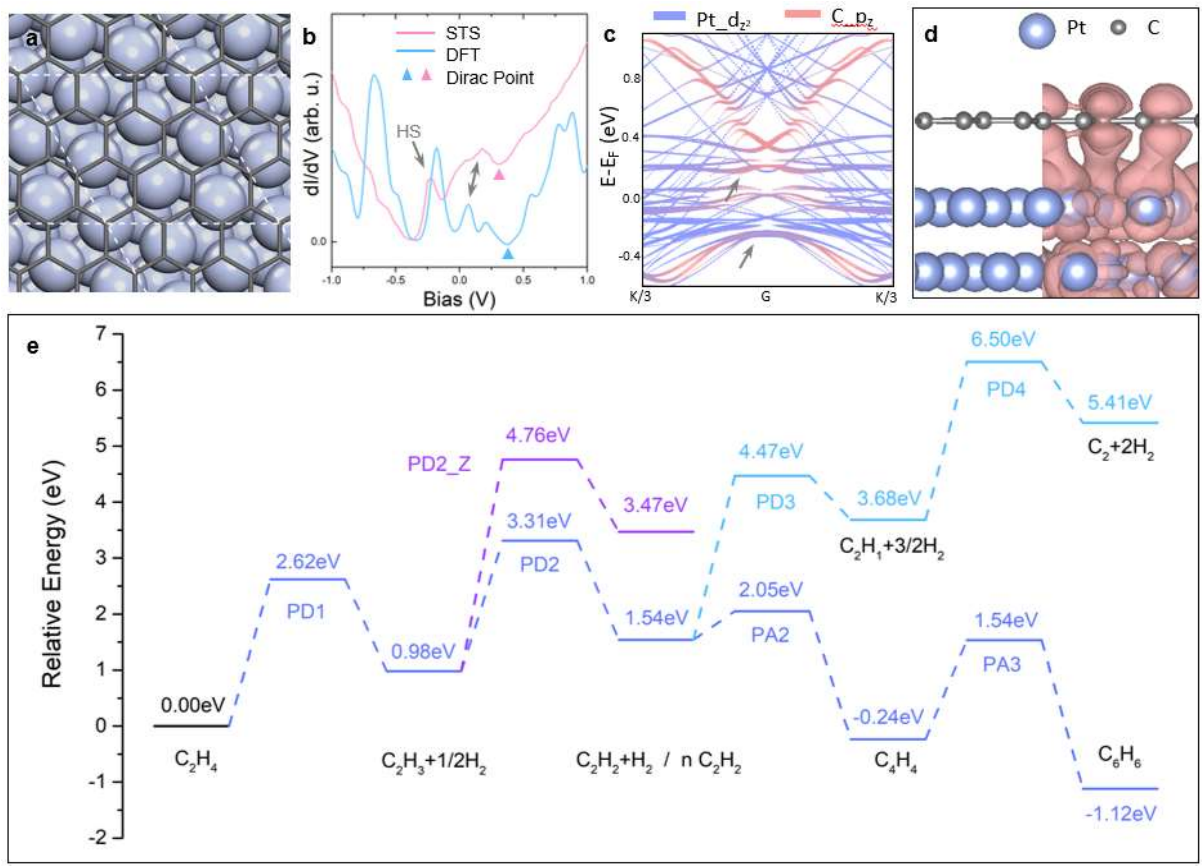

**Extended Data Fig. 2 | Theoretical evidence for the FM-mode growth of BLG on MLG-(3x3)/Pt(111)-(√7x√7)-R19°. a**, Atomic structure of the MLG/Pt(111) supercell. **b**, Experimental STS spectrum (red) and corresponding theoretical plot (blue) of the density of states of the supercell. The Dirac point (DP) and C-Pt hybridised states (HS) were highlighted by the triangles and arrows, respectively. **c**, Projected band structures of the supercell with orbital decomposition of Pt $d_{z2}$ (blue) and C $p_z$ (red). Those C-Pt hybridised bands show a mixture of red and blue colours. **d**, Visualised wavefunction norm square of a hybridised state residing at -0.05 eV and the Gamma point. The isosurface value of the contours is $2 \times 10^{-4}$ $e$/Bohr$^3$. **e**, Three potential pathways and their corresponding barrier heights for the $C_2H_4$ dissociation reaction and subsequent radical nucleation on the MLG/Pt(111) surface. Each step is denoted in the same scheme as Extended Data Fig. 1
25 / 32

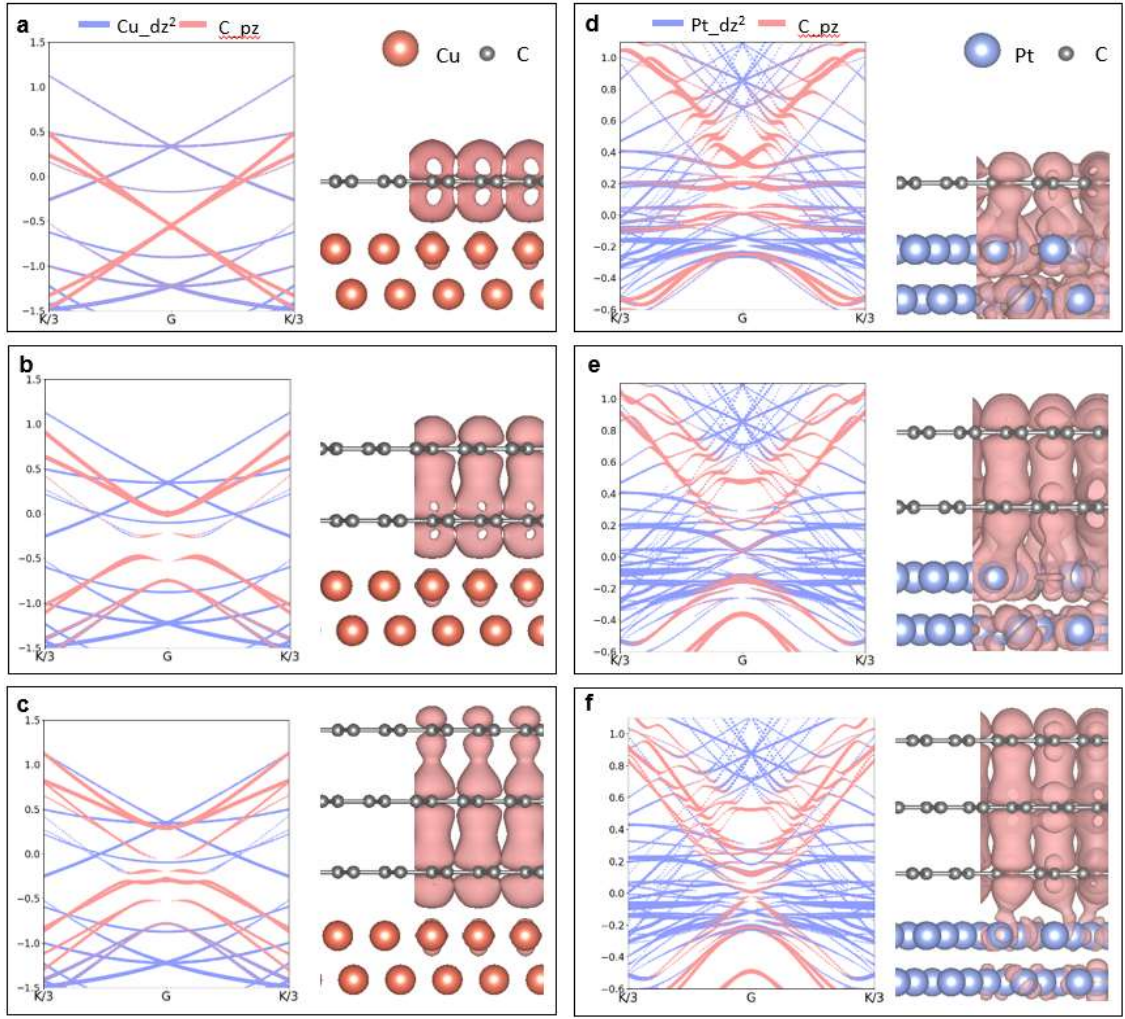

**Extended Data Fig. 3 | Electronic shybridisation between graphene and metal (Pt/Cu) surfaces.** Projected band structures with orbital decomposition of Cu(Pt) $d_{z2}$ (blue) and C $p_z$ (red) and visualised square of wavefunction norms of representative potential hybridised states of 1L (a, d), 2L (b,e) and 3L (c,f) graphene on Cu(111) (a-c) and Pt(111) (d-f), respectively. The isosurface values for all contours are $2\times10^{-4}$ $e$/Bohr$^3$. Those C-Cu(Pt) shybridised bands show a mixture of red and blue colours. A weak electronic hybridisation between carbon and copper is evidenced by relatively rare overlaps of the C $p_z$ and Cu $d_{z2}$ orbitals and the small population of the wavefunction norms on Cu atoms. In terms of FLG/Pt(111) interfaces, stronger carbon-platinum electronic hybridisation was clearly indicated from those bands showing a mixture of blue and red colours and the population of those visualised wavefunction norms on both graphene layers and Pt surfaces.

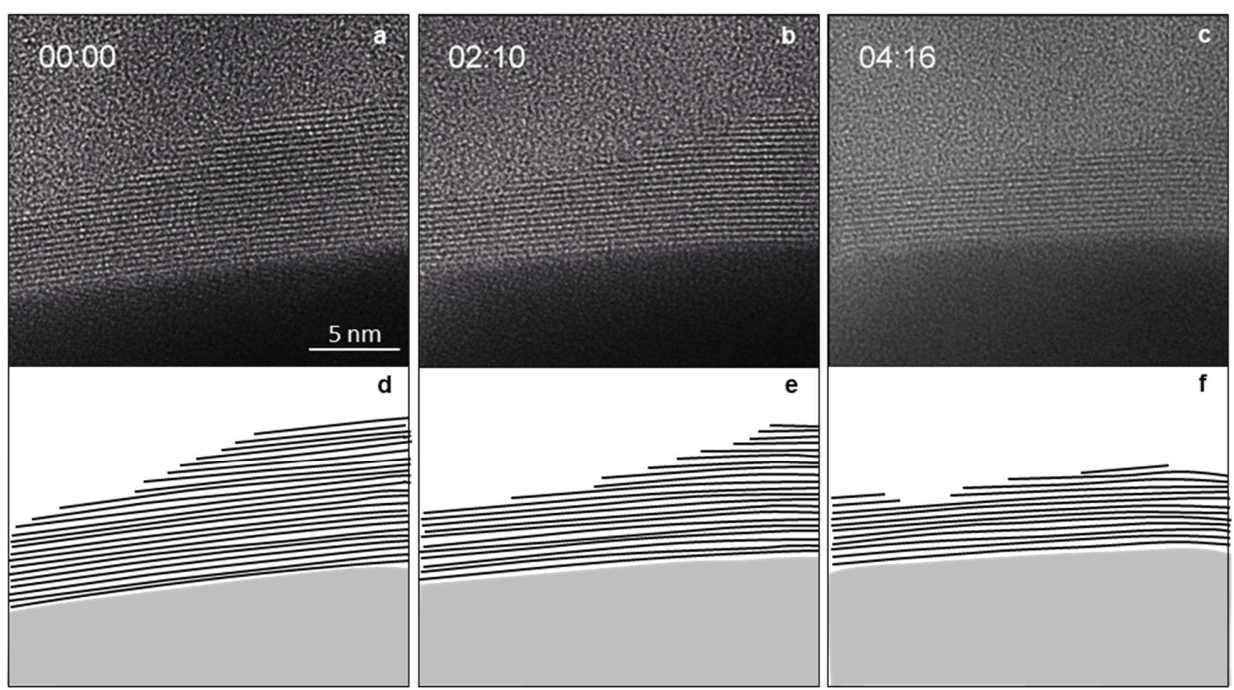

**Extended Data Fig. 4 | *In-situ* cross-sectional TEM observation recorded at 900 °C. a-f,** *In-situ* TEM images (a-c) and their associated schematics (d-f) of an etching process for FLG. It clearly shows a thinning process of graphene ad-layers on the Pt surface at different etching times using an elevated hydrogen pressure of 1000 Pa.

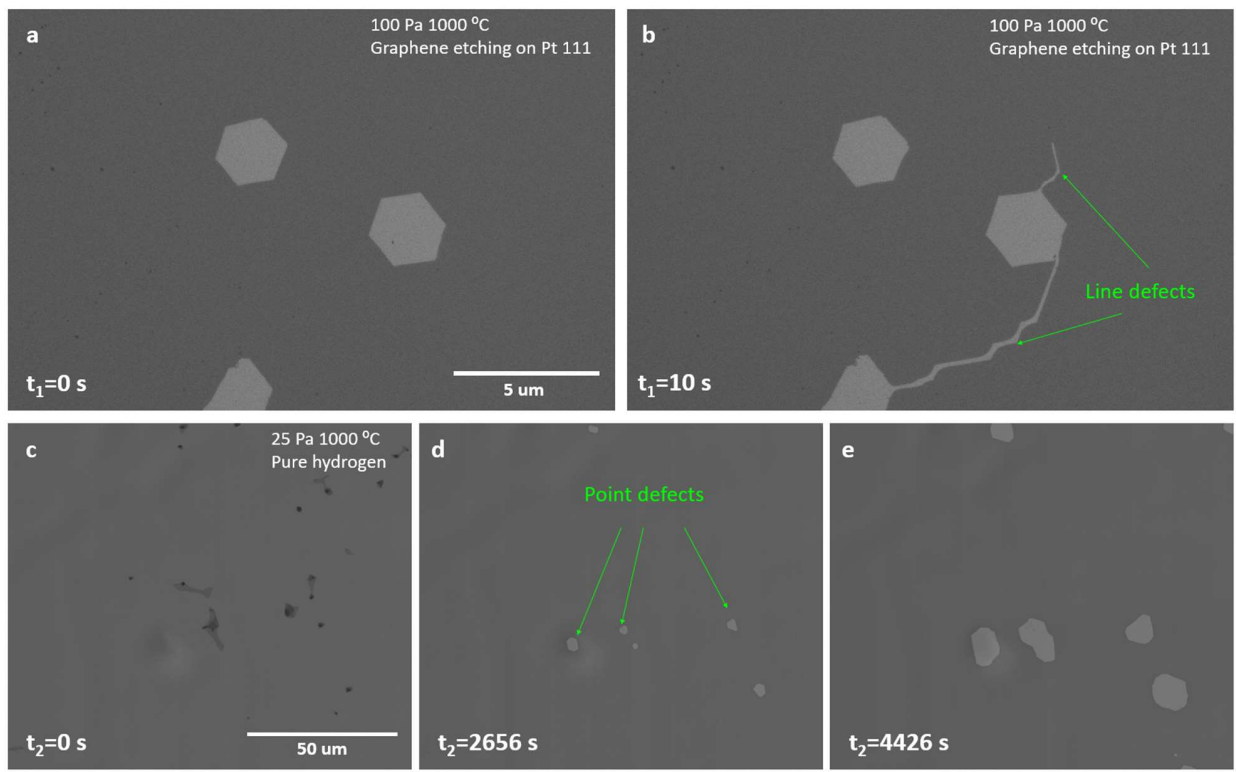

**Extended Data Fig. 5 | Time-lapse image series recorded during etching of BLG on Pt(111). a,b,** Etching of line defects (grain boundaries) in a defective bilayer graphene (BLG) film at 1000 ℃ under 100 Pa of pure hydrogen gas. Etching lines are formed when grain boundaries are present in the graphene film. **c-e,** Etching of point defects in a healed BLG film at 1000 ℃ under 25 Pa of pure hydrogen gas. The etching of point defects leads to the formation of hexagonal holes only, suggesting BLG free of line defects, in other words, domain boundaries.

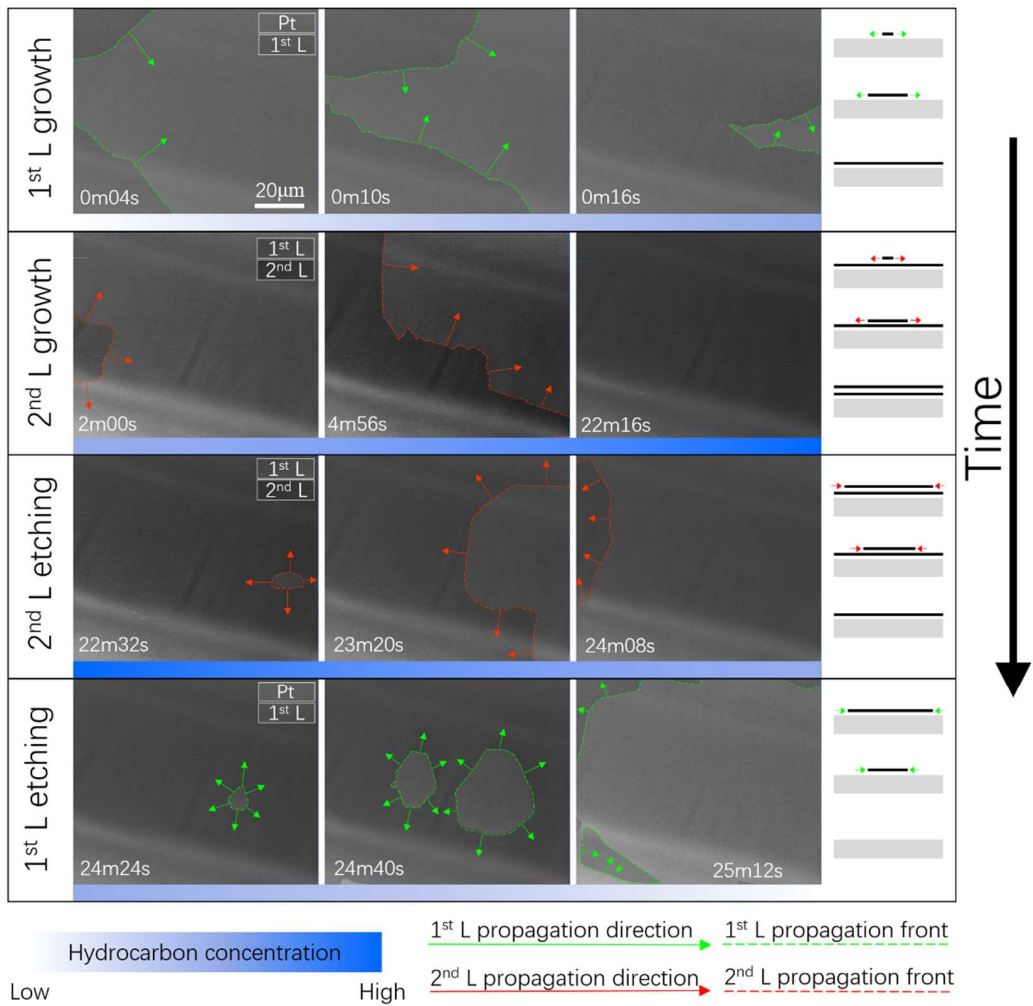

**Extended Data Fig. 6 |** One-round continuous *in-situ* SEM image sequence showing a comprehensive picture of the dynamics of the FM mode growth, healing and etching processes of BLG in UHV. The image sequence was cut from Supplementary Movie 4 recorded in a UHV-SEM. For growth, the temperature was slightly decreased to allow nucleation of the 1st and 2nd layers, while slightly increased for etching.

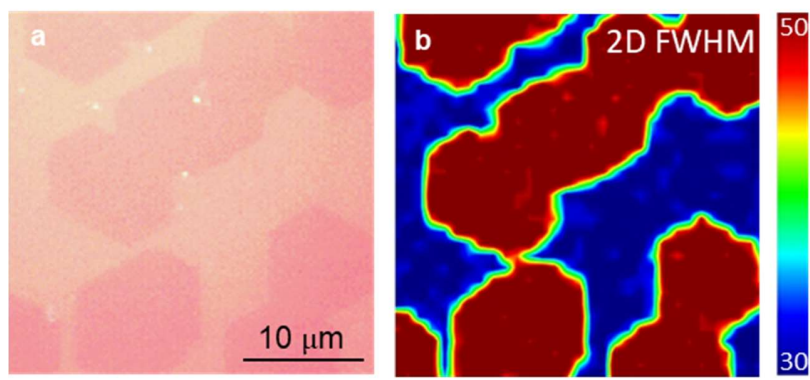

**Extended Data Fig. 7 | Optical and Raman characterisations of BLG domains prepared using tube-furnace CVD on MLG/Pt(111) .** Optical image (a) and Raman FWHM mapping of the 2D peak (b) of aligned AB-stacked BLG after being transferred onto a $SiO_2$/Si substrate.

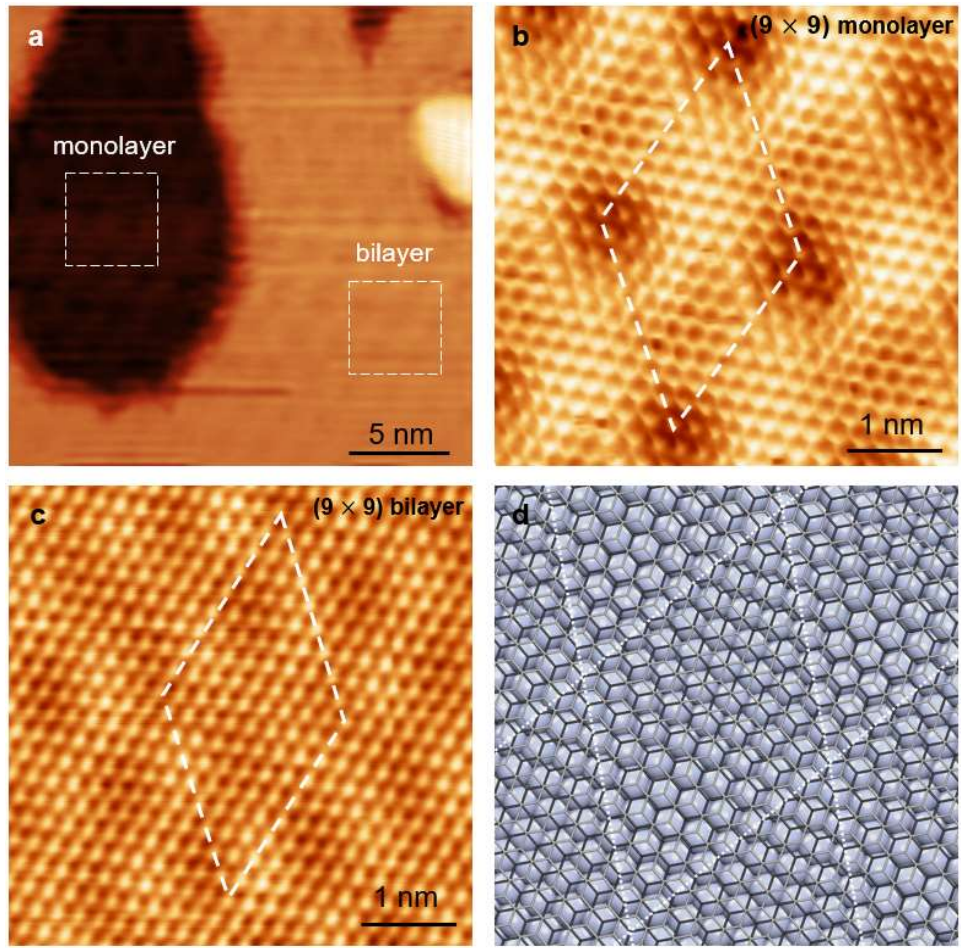

**Extended Data Fig. 8 | STM characterisation of MLG- and BLG -(9x9)/Pt(111)-(8x8)-R0°. a,** Large-scale STM image (1 nA, -100 mV). **b-c,** Atomic-resolution STM images of MLG (5 nA, -100 mV) (**b**) and BLG (100 pA, -1 V) (**c**) on Pt(111) with a rotation angle of 0°. The Moiré superlattice is marked using the white dashed lines. **d**, Fully relaxed atomic structure model of the BLG/Pt(111)-R0° superlattice where the white dotted lines denote the Moiré superlattice.

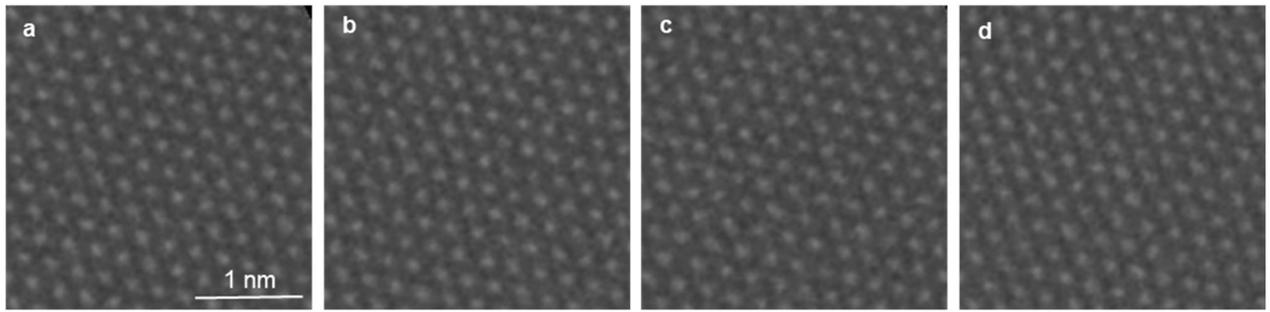

**Extended Data Fig. 9 | STEM images of BLG.** All those four panels were acquired at different locations of the tube-furnace CVD grown sample. It clearly shows AB-stacking.